\documentclass{aa}  
\usepackage{graphicx}
\usepackage{txfonts}
\begin{document}


\title{The stochastic background of gravitational waves due to the $f$-mode instability in neutron stars}

  \author{M. Surace
   \inst{1,2}
          \and
          K. D. Kokkotas 
          \inst{1}
          \and
          P. Pnigouras
          \inst{1}
          }
            
\institute{ 
             \inst{} Theoretical Astrophysics, IAAT, Eberhard-Karls University of T\"ubingen, 72076 T\"ubingen, Germany\\
               	\and
              \inst{} Institute of Cosmology and Gravitation, University of Portsmouth, Burnaby Rd, Portsmouth, PO1 3FX, UK\\
              \email{marco.surace@port.ac.uk}
      \\
             }
 
\date{\today}

\abstract
{This paper presents an estimate for the spectral properties of the stochastic background of gravitational waves emitted by a population of hot, young, rapidly rotating neutron stars throughout the Universe undergoing $f$-mode instabilities, formed through either core-collapse supernova explosions or the merger of binary neutron star systems. Their formation rate, from which the gravitational wave event rate is obtained, is deduced from observation-based determinations of the cosmic star formation rate. The gravitational wave emission occurs during the spin-down phase of the $f$-mode instability. For low magnetized neutron stars and assuming 10\% of supernova events lead to $f$-mode unstable neutron stars, the background from supernova-derived neutron stars peaks at $\Omega_{\text{gw}} \sim 10^{-9}$ for the $l=m=2$ $f$-mode, which should be detectable by cross-correlating a pair of second generation interferometers (e.g. Advanced LIGO/Virgo) with an upper estimate for the signal-to-noise ratio of $\approx$ 9.8. The background from supramassive neutron stars formed from binary mergers peaks at $\Omega_{\text{gw}} \sim 10^{-10}$ and should not be detectable, even with third generation interferometers (e.g. Einstein Telescope).}


\maketitle


\section{Introduction} \label{sec:}

The era of gravitational wave (GW) astronomy is almost upon us with the arrival of second generation terrestrial laser interferometers such as Advanced LIGO (aLIGO) and Advanced Virgo (aVirgo), with the former in operation as of 2015. Among the potentially detectable sources of GWs, $f$-mode unstable neutron stars (NS) are promising detection targets due to the copious amounts of GWs emitted, with an energy of up to $\approx 2.5\%$ of a solar mass emitted in GWs as the initially rapidly rotating NS spins down.

Non-radial, non-axisymmetric oscillations, or modes, can be excited in proto-NSs after their formation in a core-collapse supernova explosion (CCSNe). These modes may become unstable due to the GW-driven secular Chandrasekhar-Friedman-Schutz (CFS) instability \citep{PhysRevLett.24.611,1978ApJ...222..281F}, and hence grow in amplitude to produce detectable GWs. The CFS instability occurs when a star is rotating sufficiently rapidly that a retrograde mode, in the corotating frame, is dragged along with the star's rotation and appears prograde in the inertial frame.The mode angular momentum becomes increasingly negative, and hence it increases in amplitude, due to the emission of GWs which drives the mode unstable. Once the mode amplitude has saturated the star spins down, losing rotational energy via GW emission.

Another possible source of CFS-unstable modes is supramassive NSs (i.e. exceeding the maximum mass of the non-rotating star) formed through the merger of a binary NS system. Such stars have been proposed recently as an explanation for the X-ray afterglow plateau observed in some short gamma-ray bursts (GRBs) \citep{RowlinsonEtAl2013, HotokezakaEtAl2013, LaskyEtAl2014, DallOssoEtAl2015} and have lifetimes that can reach up to $\approx 4\times 10^4$ s \citep{RaviLasky2014}. Provided that the mass of these objects is below a certain threshold so that they do not promptly collapse to a black hole \citep{HotokezakaEtAl2013}, it has been shown recently that the CFS instability will grow quickly in them within $\sim10-100$ s \citep{PhysRevD.92.104040}.

Among the modes which can be driven unstable, the $f$-modes (fundamental pressure modes) and the $r$-modes (inertial modes) are particularly interesting due to their relatively short growth time scale and their efficient emission of GWs. A recent estimate for a relativistic NS has shown that the GWs from $f$-mode unstable supernova-derived NSs could be detected by aLIGO/aVirgo up to a distance corresponding to that of the Virgo Cluster ($\sim$ 15 Mpc) for reasonably high saturation amplitude values \citep{PhysRevD.87.084010}, whereas the GWs from supramassive merger-derived NSs could be detected by aLIGO/aVirgo up to a distance of 20 Mpc or by the proposed third generation detector Einstein Telescope (ET) up to 200 Mpc  \citep{PhysRevD.92.104040}. The detection of $f$-modes from individual stars could be used to perform GW asteroseismology, enabling the study of NS interiors \citep{AnderssonKokkotas1996, AnderssonKokkotas1998, KokkotasEtAl2001, GaertigKokkotas2011}.

The superposition of unresolved and uncorrelated GW signals from many sources throughout the Universe results in a stochastic background of gravitational waves (SGWB). Many contributions to the astrophysical SGWB have been discussed in the literature (see e.g. \cite{1674-4527-11-4-001} for a review), including the background due to $r$-mode instabilities in NSs. In this work, following a procedure first presented in \cite{PhysRevD.58.084020} and then in \cite{Ferrari21021999}, we provide the first estimate of the SGWB due to the $f$-mode instability in hot, young, rapidly rotating NSs formed from either a CCSNe or the merger of binary NSs, using a relativistic NS model, a realistic equation of state (EoS), and observationally-derived fits to the evolution of the cosmic star formation rate (CSFR).

The detection of an astrophysical SGWB can improve our understanding of the sources and help constrain their properties, for example neutron star ellipticities, or the average chirp mass and coalescence rate of compact binaries \citep{PhysRevD.89.123008, PhysRevLett.109.171102}. Moreover, the astrophysical SGWB will likely mask a primordial SGWB in the frequency band of terrestrial GW detectors produced during the early Universe, which if detected would shed light on the Universe shortly after the Big Bang \citep{0004-637X-659-2-918, Maggiore2000283}. Therefore, a proper understanding of the astrophysical SGWB is required in order to identify the primordial one.

This paper is organised as follows: in Section II the NS formation rate from supernovae is calculated using the fits to the CSFR; in Section III the NS formation rate this time from binary mergers is calculated; in Section IV the evolution of the $f$-mode instability is described; in Section V the spectral properties of the stochastic background are investigated; in Section VI the results are presented, namely the integrated flux and dimensionless energy density of the background; in Section VII the signal-to-noise ratio of the background using pairs of GW detectors is calculated; finally, the conclusions are drawn in Section VIII.

Note that throughout this paper the so-called 737 $\Lambda$CDM cosmology is assumed with Hubble constant $H_0=100 h_0$ km s$^{-1}$ Mpc$^{-1}$ where $h_0=0.7$, and density parameters $\Omega_m = 0.3$, $\Omega_\Lambda=0.7$.

\section{The rate of NS formation from core-collapse supernova explosions}

The rate of NS formation throughout the Universe depends upon the rate of star formation, since massive stars which end their lives in a CCSNe are the progenitors of NSs. The NS formation rate determines the GW event rate, and hence the spectral properties of the SGWB.

For the case of supernova-derived NSs, the number of NSs formed per unit time within the comoving volume out to redshift $z$ is given by \citep{Ferrari21021999}
\begin{equation} R_{\text{NS}}(z)=\int^z_0 \dot{\rho}_\star(z') \frac{dV}{dz'}dz' \int^{25M_\odot}_{8M_\odot} \Phi(m) dm \,. \end{equation}
Here, $\dot{\rho}_\star(z)$ is the CSFR (in $M_\odot \, \text{yr}^{-1} \text{Mpc}^{-3}$), i.e. the mass of gas that is converted into stars per unit time in the observer frame per unit comoving volume, $\Phi(m)$ is the stellar initial mass function (IMF) (in $M_\odot^{-2}$), i.e. the initial mass distribution of stars at the time of their birth, and $dV/dz$ is the comoving volume element. Following \cite{Ferrari21021999}, we assume that stars with masses between 8 and 25 $M_\odot$ give rise to NSs, and we adopt a standard Salpeter IMF of the form $\Phi(m) \propto m^{-(1+x)}$ with $x=1.35$, normalised through the relation
\begin{equation} \int^{125 M_\odot}_{0.1M_\odot}m\Phi(m)dm=1 \,. \end{equation}
The comoving volume element is given by
\begin{equation}\frac{dV}{dz}=4\pi\frac{c}{H_0}\frac{r(z)^2}{E(z)}, \end{equation}
where $c$ is the speed of light, $E(z)=\sqrt{\Omega_\Lambda+\Omega_m(1+z)^3}$, and $r(z)$ is the comoving distance, given by
\begin{equation} r(z)=\frac{c}{H_0}\int^z_0\frac{dz'}{E(z')}\,, \end{equation}
assuming spatial flatness, i.e. $\Omega_m+\Omega_\Lambda=1$.

Since the CSFR is not a directly observable quantity, it is usually inferred from observations of the rest-frame ultraviolet (UV) light, as it is mainly radiated by short-lived massive stars. The UV galaxy luminosity density is studied by space and ground-based telescopes, and is then converted into the CSFR density through the adoption of a universal stellar IMF to calculate the conversion factor. Other authors have made use of observations of the GRB rate at high redshift to infer the CSFR. Note that the use of a standard Salpeter IMF does not introduce considerable errors because the evaluation of the NS formation rate based on the CSFR and an assumed universal IMF is largely independent of the specific IMF adopted \citep{1998ASPC..146..289M}.

As there exist many parameterized fits to the expected evolution of the CSFR with redshift in the literature, we adopt four recent ones to account for the uncertainties: HB06 \citep{0004-637X-651-1-142}, F07 \citep{Fardal11082007}, W08 \citep{Wilkins01042008}, based on UV observations; and RE12 \citep{0004-637X-744-2-95}, based on the observed GRB rate. The CSFR fits, plotted in Fig. 1, rise rapidly from their local values to peak between $z \sim 1.5 - 2.5$ for the UV derived fits, and much later at $z \sim 4$ for the GRB derived fit, falling again at higher redshifts. Up to $z \sim 0.5$ all three UV derived fits are in close agreement, while at increasing redshift the fits diverge. The cutoff for each curve corresponds to the maximum redshift of each CSFR fit: $z=6$ for the UV derived fits, and $z=15$ for the GRB derived fit.

The uncertainty on the CSFR at high redshift does not present a serious problem because the contributions to the SGWB come mainly from low redshift sources. The energy flux emitted by a single source decreases as the inverse square of the luminosity distance, so high redshift sources provide a negligible contribution to the background.

 \begin{figure}[]
\begin{center}
\includegraphics[width=0.5\textwidth]{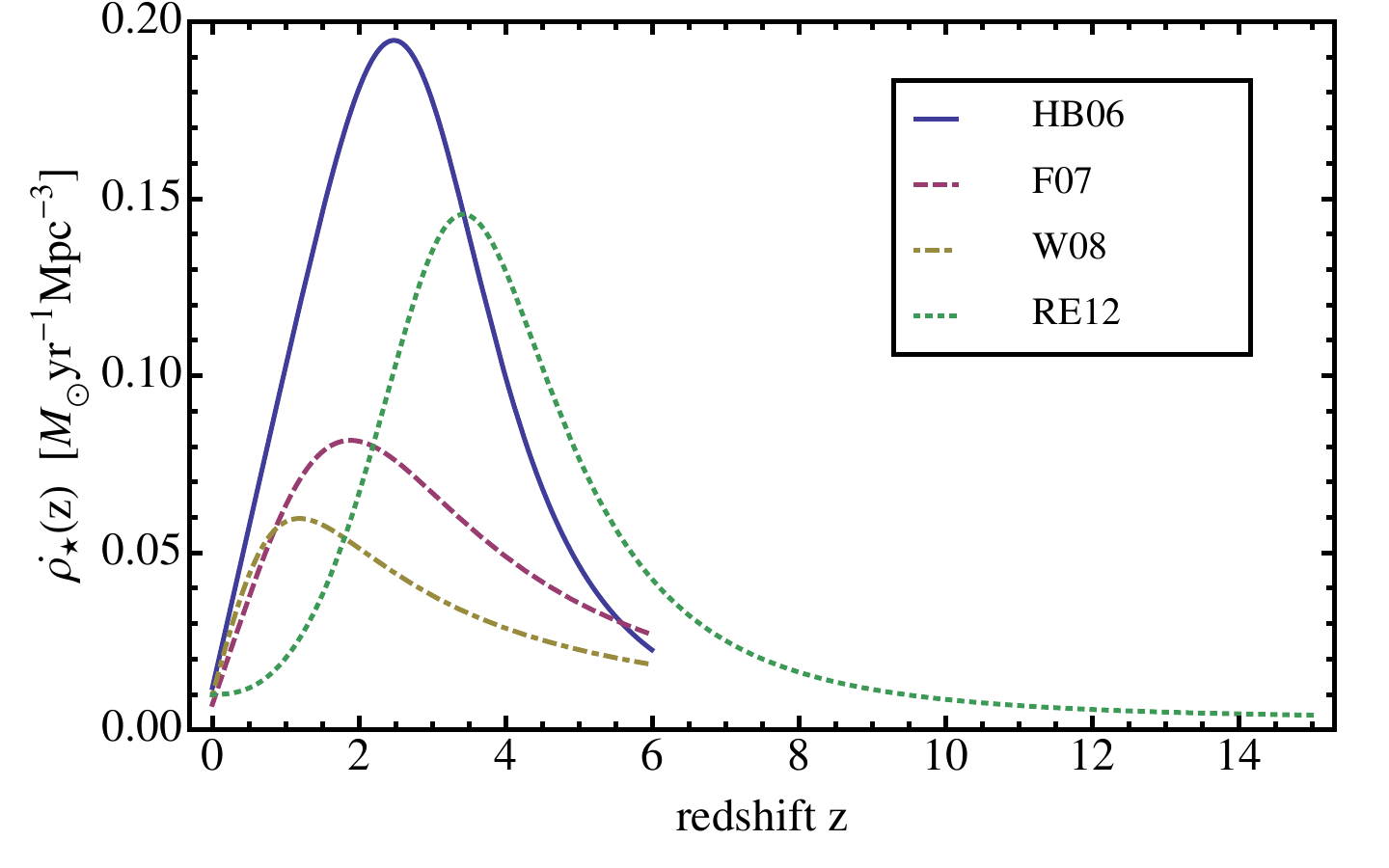}
\caption{Evolution of the cosmic star formation rate density with redshift for four different CSFR fits (see Section II).}
\label{default}
\end{center}
\end{figure}

The NS formation rate defined in eq. (1) is plotted in Fig. 2. The HB06 fit  gives the highest $R_\text{NS}$ as expected since it provides the highest CSFR up to $z \sim 4$, where the majority of star formation takes place. At high redshift, $R_{\text{NS}}$ plateaus in accordance with the decreasing CSFR. Note that the RE12 CSFR fit gives almost the same $R_{\text{NS}}$ at low redshifts up to $z\sim 0.5$, although its behaviour differs significantly from the UV-derived fits (even at low redshifts).

\section{The rate of NS formation from the merger of NS binaries}

 \begin{figure}[]
\begin{center}
\includegraphics[width=0.5\textwidth]{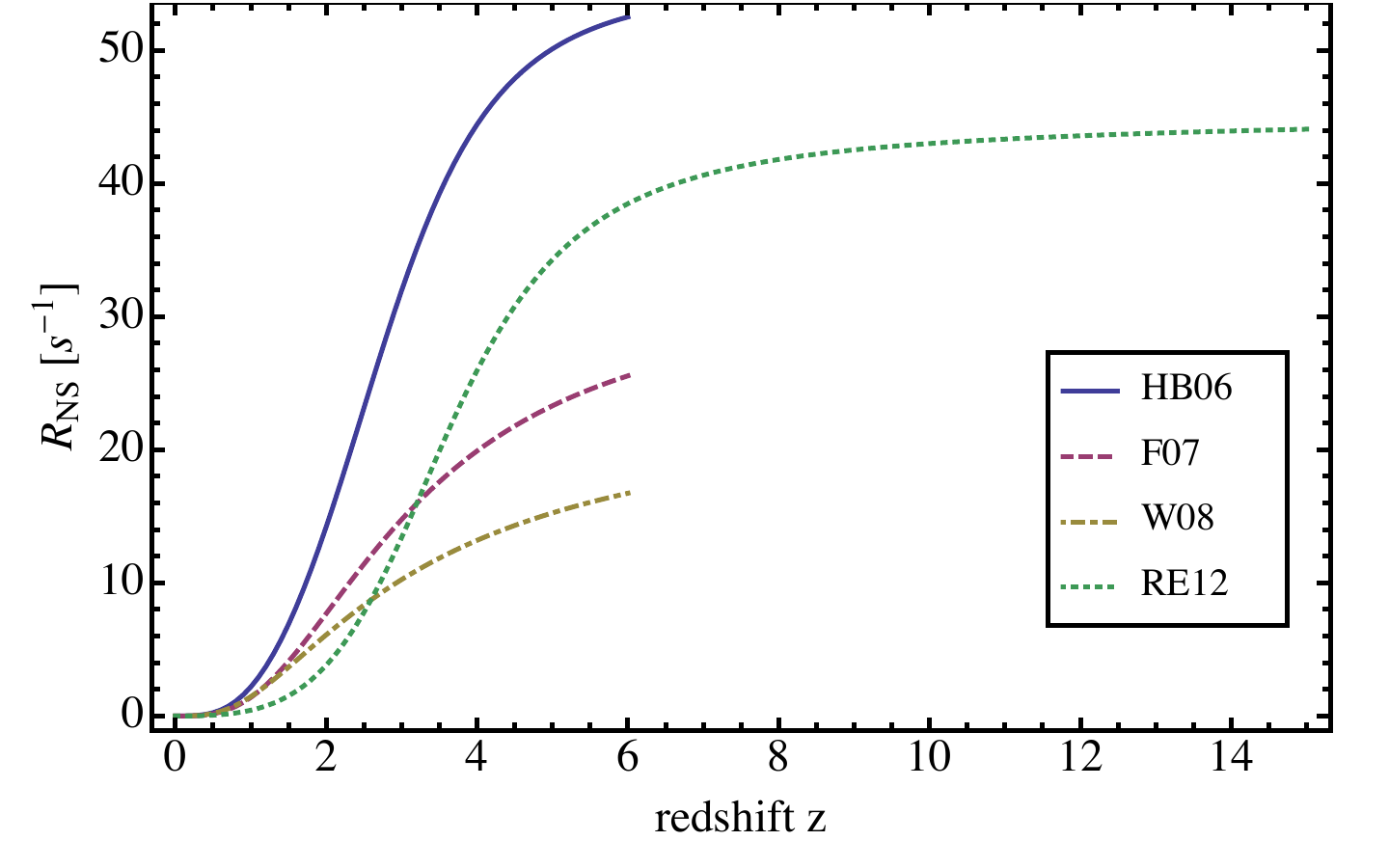}
\caption{Evolution of the number of NSs formed per unit time within the comoving volume out to redshift $z$ based on four different CSFR fits (see Section II), for NSs formed from a CCSNe.}
\label{default}
\end{center}
\end{figure}

For NSs formed through the merger of binary NS systems, we assume that the merger rate tracks the star formation rate with some delay $t_d$ from formation of the binary to final merger. Then the observed rate of binary coalescence (mergers) per unit volume at redshift $z$ is given by \citep{PhysRevD.79.062002}
\begin{equation} \dot{\rho}^o_c(z)= \dot{\rho}^o_c(0) \times \frac{\dot{\rho}_{\star,c}(z)}{\dot{\rho}_{\star,c}(0)} \,,\end{equation}
where $\dot{\rho}^o_c(0)$ is the rate in the local Universe, and $\dot{\rho}^o_c(z)$ is normalized to reproduce the local rate for $z=0$. The local merger rate $\dot{\rho}^o_c(0)$ is usually extrapolated by multiplying the rate in the Milky Way with the density of Milky Way galaxies. Following \cite{PhysRevD.79.062002}, we take $\dot{\rho}^o_c(z=0) \sim 1 \, \text{Myr}^{-1} \text{Mpc}^{-3}$ as its most probable value.

The quantity $\dot{\rho}_{\star,c}(z)$ relates the past star formation rate to the rate of binary merger and is defined as
\begin{equation} \dot{\rho}_{\star,c}(z) = \int \frac{\dot{\rho}_{\star}(z_f)}{(1+z_f)}P(t_d)dt_d \,,  \end{equation}
where $\dot{\rho}_\star$ is the CSFR introduced above. Here, $z$ describes the redshift when a compact binary merges, and $z_f$ is the redshift at which its progenitor binary formed. The time delay $t_d$ connects these two redshifts, and represents the total time from initial binary formation to evolution into a compact binary, plus the merging time $\tau_m$.  The quantity $t_d$ is also the difference in lookback times between $z_f$ and $z$,
\begin{equation} t_d=t_{\text{LB}}(z_f)-t_{\text{LB}}(z)= \frac{1}{H_0}\int \frac{dz'}{(1+z')E(z')}\,,     \end{equation}
with $E(z)$ defined above. In eq. (6), $P(t_d)$ is the probability distribution for the delay time $t_d$, and the factor $1/(1+z_f)$ accounts for time dilation due to the cosmic expansion. Population synthesis (see references in \cite{PhysRevD.79.062002}) suggests that $P(t_d)$ takes the form
\begin{equation}   P_d(t_d) \propto \frac{1}{t_d} \;\;\text{with}\;\; t_d > \tau_0 \end{equation}
for some minimal delay time $\tau_0$. Again, following \cite{PhysRevD.79.062002}, we assume $\tau_0 \sim 20$ Myr, which corresponds roughly to the time it takes for massive binaries to evolve into two NSs.

The rate of binary NS mergers within the comoving volume out to redshift $z$ is given by
\begin{equation} R_\text{BM}(z)=\int \dot{\rho}^o_c(z') \frac{dV}{dz'}dz' \,,\end{equation}
and is plotted in Fig. 3 for the four CSFR fits described above. Here, it is apparent that NSs formed from binary mergers are far less numerous throughout the Universe than supernova-derived NSs, since $R_\text{BM}$ is roughly an order of magnitude lower than $R_\text{NS}$. The curves are in agreement up to $z \sim 1$, a higher $z$ than in Fig. 2, and now the RE12 fit provides the highest $R_\text{BM}$ for $z>3$, even though it gives the highest CSFR only for $z>4$. This is due to the normalisation of $\dot{\rho}^o_c(z)$, so that the local value at $z=0$ is the same for all the CSFR fits, which dampens the effect of a high CSFR at low redshifts.

 \begin{figure}[]
\begin{center}
\includegraphics[width=0.5\textwidth]{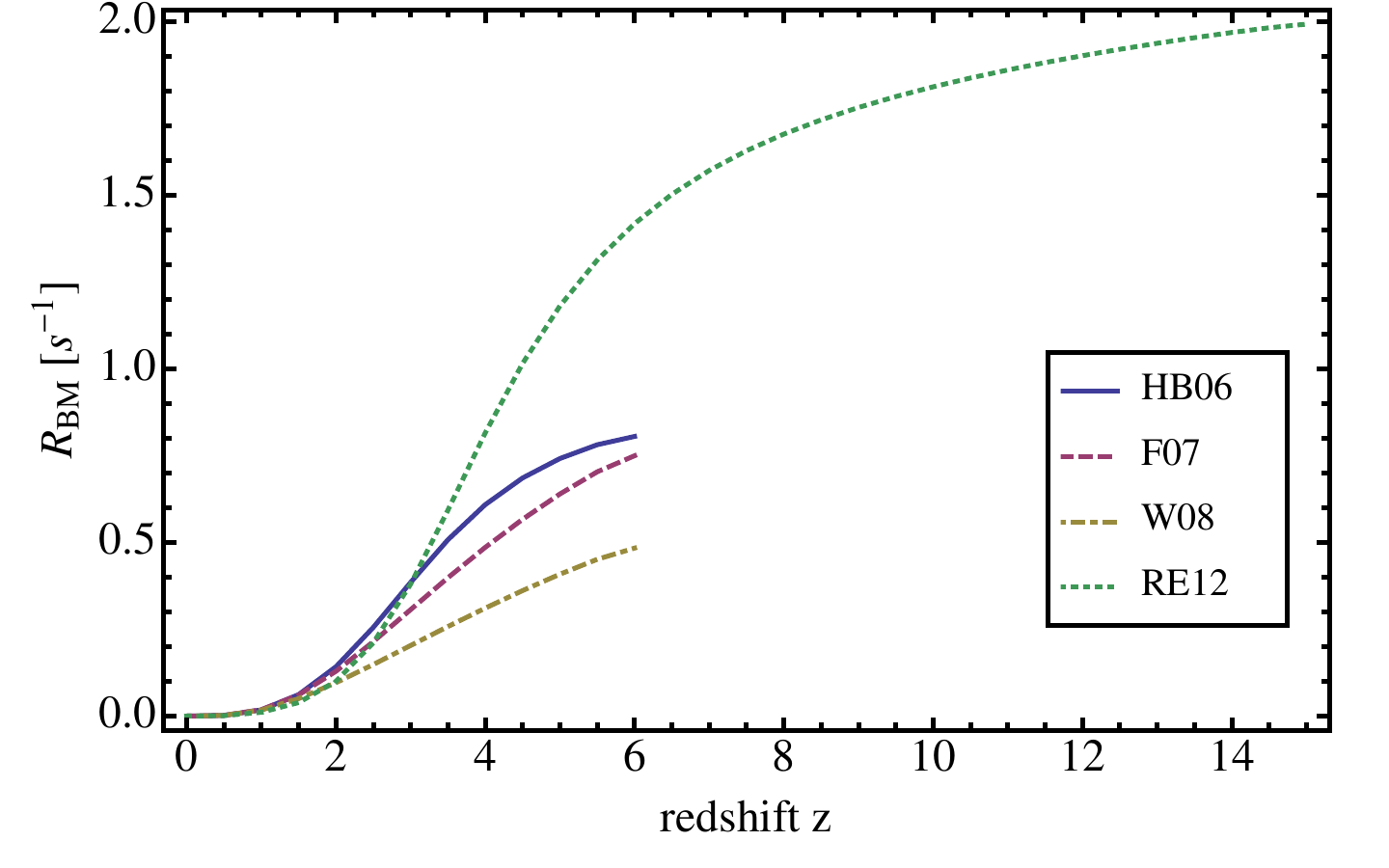}
\caption{Evolution of the number of NS binary mergers per unit time within the comoving volume out to redshift $z$ based on four different CSFR fits (see Section II).}
\label{default}
\end{center}
\end{figure}

\section{The evolution of the $f$-mode instability}

For $f$-mode unstable NSs, the evolution of the instability proceeds as follows. Initially, the NS is very hot ($\sim 10^{11}$ K) and rotates very rapidly at close to its Kepler limit, i.e. the angular velocity $\Omega$ at which mass shedding at the equator occurs, making the star unstable. Note that although the distribution of initial rotation periods for supernova-derived NSs is uncertain, as a conservative guess we assume that 10$\%$ of the population rotates initially at close to the Kepler limit, with the remainder rotating too slowly to become $f$-mode unstable. For NSs formed from binary mergers, all of the stars are expected to initially rotate at the Kepler limit. We also assume that both NS classes rotate uniformly, since differential rotation is expected to disappear shortly after the star's formation \citep[and references therein]{PhysRevD.92.104040}.

The amplitude of the $f$-mode is initially negligible but grows exponentially due to the CFS instability. Meanwhile, the NS cools due to neutrino emission. Eventually, the amplitude of the mode saturates due to nonlinear coupling with other modes in the star which drain the energy of the $f$-mode, after which it remains roughly constant. The star then spins down via GW emission at approximately constant temperature ($T \sim 10^9$ K) as the heat generated by shear viscosity balances the neutrino cooling. The star stops emitting GWs when it is no longer spinning sufficiently quickly to maintain the CFS instability, and the $f$-mode is rapidly damped. 

Note that the $f$-mode saturation amplitude determines how fast the NS spins down, and thus whether the associated GW emission will be detectable in terms of single events or a continuous stochastic background. Furthermore, for NSs with a strong dipole magnetic field component at the surface ($B_p \sim 10^{12}-10^{14}$ G), a significant fraction of the star's rotational energy could be lost due to magnetic braking rather than via GWs, shortening the total evolution time and reducing the strength of the GW signal.

At temperatures above or below $\sim10^9$ K, bulk and shear viscosity respectively suppress the growth of the $f$-mode. Therefore, there exists a range of star temperatures and rotation rates for which the $f$-mode is CFS unstable and overcomes dissipative processes. This is usually represented as a curve in the ($T,\Omega$) plane above which the growth timescale due to GWs is shorter than the dissipative timescales due to bulk and shear viscosity, and is known as the instability window.

The instability windows used in this work were extracted from \cite{PhysRevD.88.044052} and \cite{PhysRevD.92.104040} for supernova and merger-derived NSs res/pectively, which have been calculated for the realistic WFF2 EoS (denoted as UV14+UVII in \cite{PhysRevC.38.1010}) in the Cowling approximation, where the spacetime metric is kept fixed. Since the evolution of the $f$-mode instability, i.e. the star's trajectory through the window, is unavailable for supernova-derived stars with this EoS, we consider the evolution calculated by \cite{PhysRevD.87.084010} for a relativistic NS model and a polytropic EoS with polytropic index $N=0.62$, which closely resembles the WFF2 EoS. In this case, the star exits the parabola-shaped window close to its minimum point, so we assume that the maximum change in $\Omega$ from the Kepler frequency to the minimum point determines the total energy lost to GWs.

The instability windows for supernova-derived NSs reach down to roughly 96$\%$ and 80$\%$ of the Kepler limit for the $l=m=2$ and $l=m=3,4$ $f$-modes respectively \citep{PhysRevD.88.044052}. For merger-derived NSs only the $l=m=2,3$ $f$-mode windows have been calculated. These post-merger supramassive NSs lose $\approx 9 \%$ of their initial rotational energy before collapsing to a black hole for both modes, assuming the star's mass is below a certain threshold so that it doesn't immediately collapse to one \citep{PhysRevD.92.104040}.

Note that abandoning the Cowling approximation and including a dynamical spacetime is expected to lead to more energy lost via GWs and a stronger SGWB, since the star is secularly unstable down to lower rotation rates in full general relativity \citep{PhysRevD.81.084055}.

\section{Spectral properties of the stochastic background}

The spectral properties of a stochastic background of GWs can be characterised by the dimensionless energy density parameter $\Omega_{\text{gw}}$, defined as
\begin{equation} \Omega_{\text{gw}}(\nu_o) = \frac{1}{\rho_c}\frac{d\rho_{\text{gw}}}{d\,\text{ln}\, \nu_o} \,, \end{equation}
where $\rho_{\text{gw}}$ is the GW energy density, $\nu_o$ is the GW frequency in the observer frame, and $\rho_c=3H^2/8\pi G$ is the critical energy density required to make the Universe flat today.

For a SGWB of astrophysical origin, the dimensionless energy density can be expressed in terms of the integrated flux received on Earth $F_{\nu_o}$ at the observed frequency (in erg cm$^{-2}$ Hz$^{-1}$ s$^{-1}$) as
\begin{equation} \Omega_{\text{gw}}(\nu_o)=\frac{1}{\rho_cc^3}\nu_oF_{\nu_o}(\nu_o)\,, \end{equation}
where $F_{\nu_o}$ is defined as
\begin{equation} F_{\nu_o}(\nu_o)=\int f_{\nu_o}(\nu_o, z)\frac{dR(z)}{dz}dz\,. \end{equation}
Here, $f_{\nu_o}$ is the energy flux per unit frequency (fluence) emitted by a single source located at redshift $z$ (in erg cm$^{-2}$ Hz$^{-1}$), which is given by
\begin{equation} f_{\nu_o}(\nu_o, z)= \frac{1}{4\pi d_L(z)^2} \left(\frac{dE_{\text{gw}}}{d\nu}\right)_o(1+z)\,. \end{equation}
The luminosity distance $d_L$ is given by $d_L(z)=(1+z)r(z)$. The GW energy spectrum $dE_{\text{gw}}/d\nu$ (in erg Hz$^{-1}$) must be redshifted to the observer frame where $\nu$ is the emitted GW frequency, which is related to the observed frequency by $\nu=\nu_o(1+z)$. The second factor in eq. (12), $dR(z)/dz$, gives the number of sources per unit time per unit redshift in the observer frame, and is obtained from eq. (1) or (9) for supernova or merger-derived NSs respectively.
 
If we assume that the rotational energy lost while the star is inside the instability window is entirely due to GW emission then, following \cite{Ferrari21021999}, we can approximate the GW energy spectrum as
\begin{equation} \frac{dE_{\text{gw}}}{d\nu} \approx 2 E_\text{lost}\frac{\nu}{\nu_\text{max}^2}\,, \end{equation}
where $E_\text{lost}$ is the change in the star's rotational energy from its initial value at the Kepler frequency to the frequency at which it exits the instability window, and $\nu_\text{max}$ is the maximum emitted frequency of GWs in the source frame. Eq. (14) can be modified with an appropriate factor if a fraction of the energy is lost due to magnetic braking instead.

Substituting the relevant expressions into eq. (11) for both NS classes, we are left with an integral over $z$ to perform that provides $\Omega_{\text{gw}}(\nu_o)$. The $z$ limits depend on both the emission frequency range in the source frame, and the redshift range of the CSFR fit, where the upper $z$ limit is given by
\begin{equation}    z_\text{sup} (\nu_o) =  \begin{cases} z_\text{max} & \text{if } \nu_o < \frac{\nu_\text{max}}{(1+z_\text{max})} \\ \frac{\nu_\text{max}}{\nu_o}-1 & \text{otherwise} \,,\end{cases}  \end{equation}
and the lower $z$ limit is given by
\begin{equation}    z_\text{min} (\nu_o) =  \begin{cases} 0 & \text{if } \nu_o > \nu_\text{min} \\ \frac{\nu_\text{min}}{\nu_o}-1 & \text{otherwise} \,.\end{cases}  \end{equation}
Here $z_\text{max}$ and $z_\text{min}$ are the maximal and minimal redshifts of the CSFR fits, respectively. Thus, the spectral shape of the SGWB is characterized by a cutoff at the maximal GW emission frequency, and a maximum at a frequency which depends on the shape of both the CSFR and the energy spectrum.

The data required to evaluate eqs. (11) and (12), including the emitted GW frequencies and the rotational energy of the star at the maximum and minimum angular velocities for which the instability operates, was extracted from \cite{PhysRevD.88.044052, PhysRevD.92.104040}. 

\section{Results}

\subsection{Integrated flux}

\begin{figure}[]
\begin{center}
\includegraphics[width=0.5\textwidth]{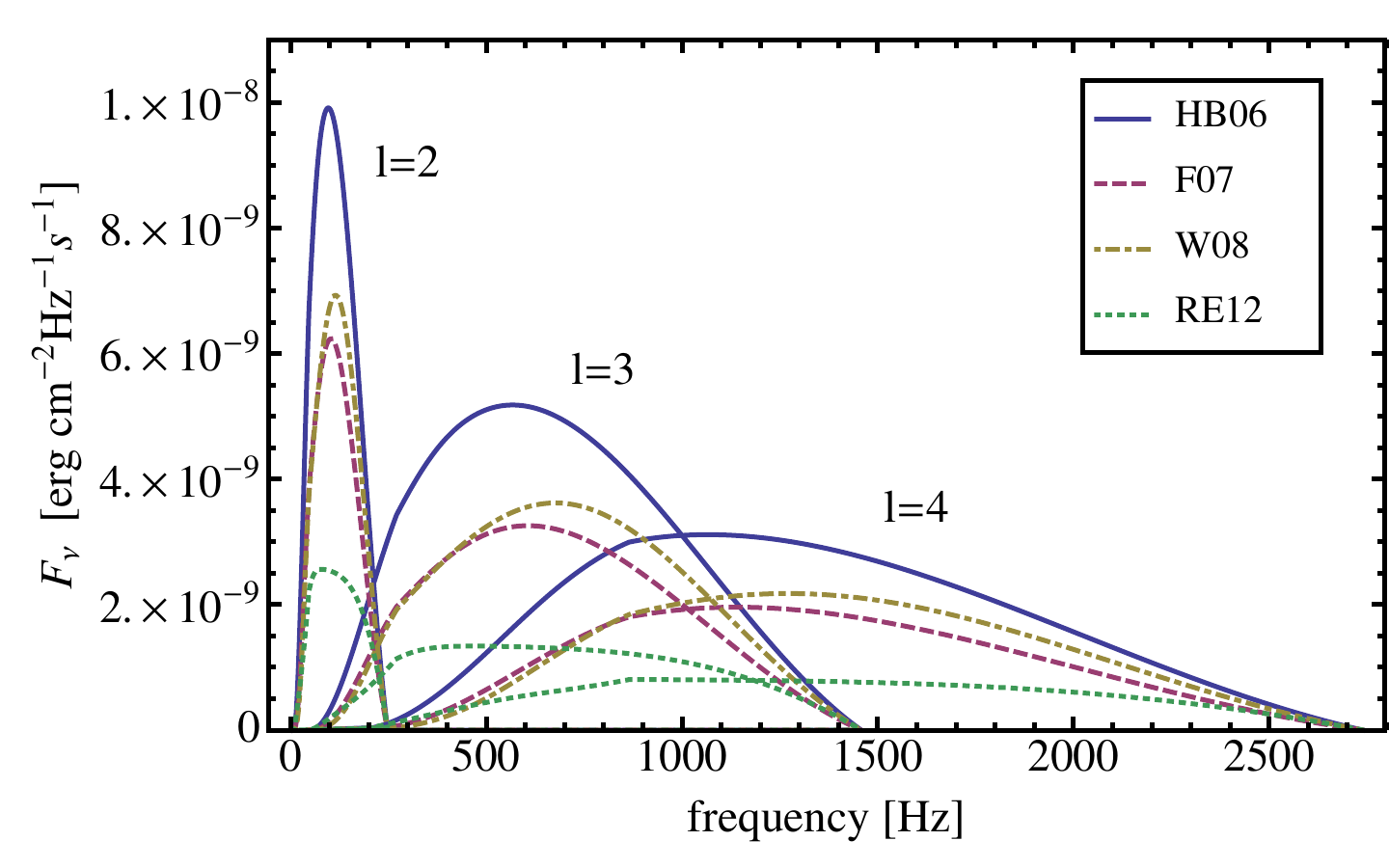}
\caption{Integrated flux of the stochastic background for supernova-derived NSs, for four CSFR fits  (see Section II) and the $l=m=2,3,4$ $f$-modes, as a function of the observed frequency.}
\label{F}
\end{center}
\end{figure}

\begin{figure}[]
\begin{center}
\includegraphics[width=0.5\textwidth]{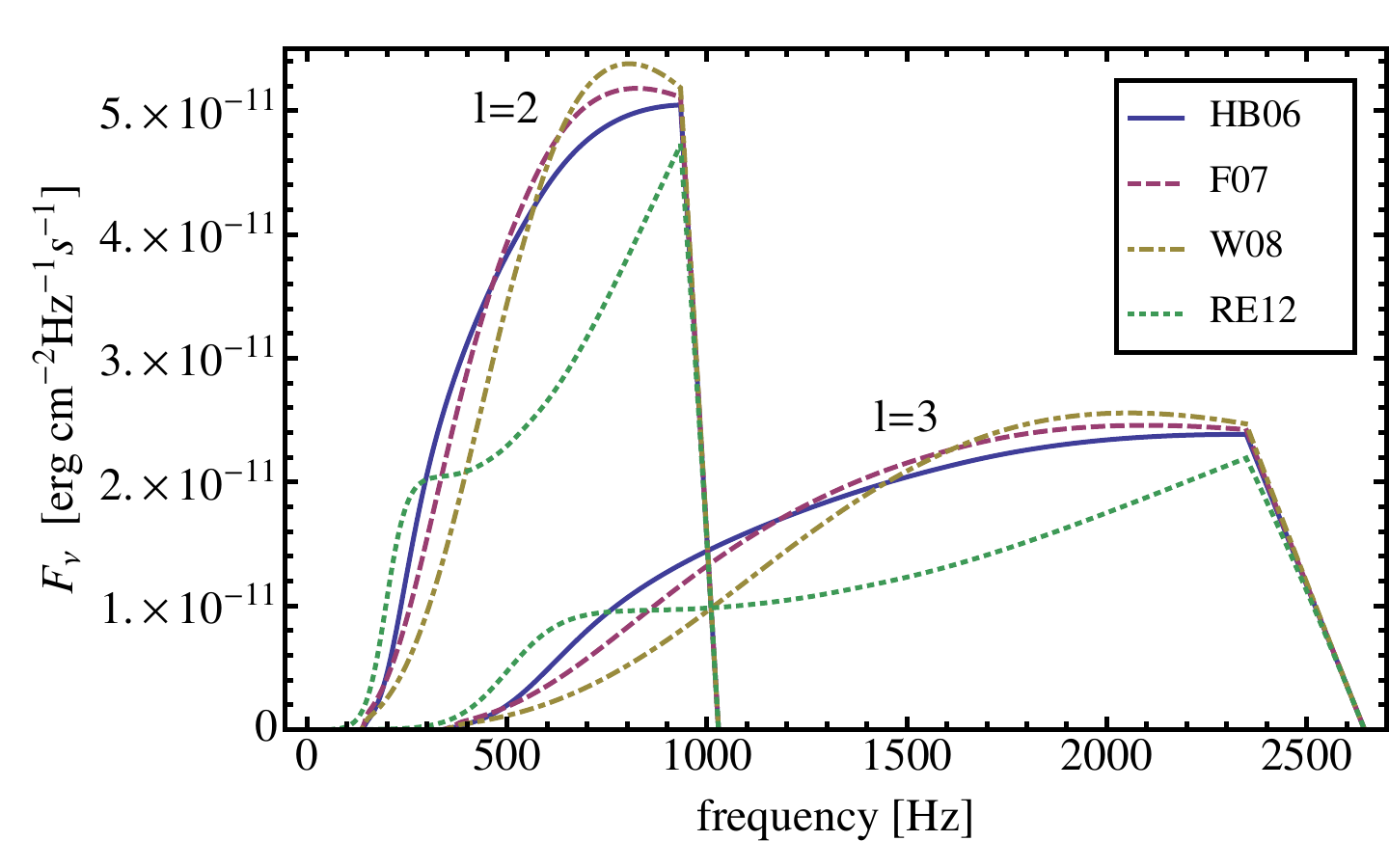}
\caption{Integrated flux of the stochastic background for binary merger-derived NSs, for four CSFR fits  (see Section II) and the $l=m=2,3$ $f$-modes, as a function of the observed frequency.}
\label{Fbin}
\end{center}
\end{figure}

The integrated flux of the SGWB for supernova and merger-derived NSs is plotted in Fig. \ref{F} and \ref{Fbin} respectively, for the four CSFR fits described in Section II. The total energy emitted in GWs during the spin-down, i.e. the area under the curves, is consistent with the rotational energy lost or the change in $\Omega$ inside the instability windows, quoted in Section IV.

The shape of the curves can be understood as follows. For a single source, a certain range of frequencies is emitted by each mode in the source frame, and the signal is stronger at higher frequencies, according to eq. (14). The spectral shape of the signal (in the source frame) would thus be wedge-shaped, with the highest flux occurring at the highest emitted frequency. As the contribution from sources at increasing redshifts is included, the same spectrum but shifted to increasingly lower frequencies is added to the signal. Thus, the observed signal over the emitted frequency range rises, with a drop at lower frequencies which becomes less steep as more sources are added. The contribution from sources at increasing redshifts also gets stronger since the CSFR rises, until it reaches a peak and drops, which decreases the contribution from new sources. Furthermore, the flux from high redshift sources is reduced according to the inverse square law.

In Fig. 4, these effects combine to produce bell-shaped curves, where the curved peaks are due to sources at the redshift where the CSFRs in Fig. 1 are highest. The HB06 curves have the highest amplitude since $R_\text{NS}$ is the highest for that fit (Fig. 2). The fact that the RE12 fit extends to higher redshifts than the other fits increases the signal power at low frequencies (compared to if the cut off was at the same $z$). Also, given that the stochastic background is mainly contributed by low redshift sources, where the $R_\text{NS}$ curves (Fig. 2) are in close agreement, the importance of the divergent CSFRs in Fig. 1 is reduced.

In Fig. 5 the curves are around two orders of magnitude weaker than in Fig. 4, since $R_\text{BM}$ is considerably lower than $R_\text{NS}$. Also, the curves are closer together in terms of flux, since $R_\text{BM}$ (Fig. 3) for the different fits is in agreement up to a higher redshift than for $R_\text{NS}$ (Fig. 2). For the RE12 and HB06 curves, the minimum emitted frequency is the highest peak rather than the curved one (which is due to sources at the redshift of maximum CSFR). This is because the emitted frequency range is far narrower and at higher frequencies for merger-derived NSs. The minimum emitted frequency is now at a higher frequency than the curved peak, and even more so for the RE12 and HB06 fits which have maximum CSFRs occurring at the highest redshifts (which shifts the curved peak to lower frequencies). This means that the curved peak does not smooth the drop in flux to the left of the minimum emitted frequency.

\subsection{Dimensionless energy density}

The dimensionless energy density of the SGWB for supernova and merger-derived NSs is plotted in Fig. 6, 7 and 8 respectively, again for the four CSFR fits described in Section II. The strongest stochastic background over $\sim1-1000$ Hz from Fig. 6 of \cite{1674-4527-11-4-001}, due to coalescing binary NSs (BNS), has been plotted on the same figures for comparison. The GW detection limits for aLIGO/aVirgo and ET have also been plotted. The sensitivity curves are taken from Table 1 of \cite{livrev}, where an analytic fitting formula is given for the noise power spectral density of each detector. The conversion to $\Omega_{\text{gw}}$ for two co-located detectors and uncorrelated instrumental noise is given in Section 8.1.2 of \cite{livrev}. 

To determine how changing our assumptions on the properties of the NS population affects the detectability of the background, a `lower limit'  has been plotted in Fig. 7. Here we have assumed that only 1\% of proto-NSs rotate sufficiently quickly to become CFS unstable, and that only half of the rotational energy is lost to GWs, with the other half lost to magnetic braking.

Assuming initially that a signal is detectable if the predicted amplitude lies above the intrinsic noise curve for that particular instrument at any frequency, then the $l=m=2$ background in Fig. 6 is the only one detectable with second generation detectors since the signal peaks at low frequencies ($\approx 50 - 200$ Hz) where the detector sensitivity is highest, and at a higher amplitude than the background due to coalescing BNS. The $l=m=3$ background might also be detectable, however it lies close to the sensitivity limit. For third generation detectors the $l=m=3$ background in Fig. 6 is detectable, however the $l=m=4$ one peaks at frequencies that are too high to be detected, even though it has the highest signal amplitude. Note, however, that the $l=m=3$ and 4 $f$-mode backgrounds peak at roughly the same frequency ($\sim 10^3$ Hz) and amplitude ($\Omega_{\text{gw}} \sim 10^{-8}$) as many other astrophysical backgrounds in Fig. 6 of \cite{1674-4527-11-4-001}, hence they might be indistinguishable from them.

The lower estimate for the $f$-mode background due to supernova-derived NSs in Fig. 7 provides a dramatically different result. Now the $f$-mode background is too weak to be detected with aLIGO/aVirgo, but the BNS background should be detectable. The $l=m=2$ background would be detectable with the ET interferometer, however it is now an order of magnitude weaker than the BNS background, so would be overpowered by it.

The background produced by merger-derived NSs in Fig. 8 should not be detectable, even for the most sensitive ET detector. This can be attributed to several reasons: firstly to the lower NS formation rate from binary mergers than from supernovae, and the fact that fewer GW sources (at low redshift) leads to a weaker accumulated GW background. Secondly, the supramassive NSs collapse to black holes before they can lose a significant proportion of their initial rotational energy, emitting less total energy in GWs, even though their individual GW emission is stronger than from supernova-derived NSs since it is detectable at greater distances. Thirdly, the background peaks at high frequencies around 1000 Hz where the interferometer sensitivity is considerably reduced.

 \begin{figure}
\begin{center}
\includegraphics[width=0.5\textwidth]{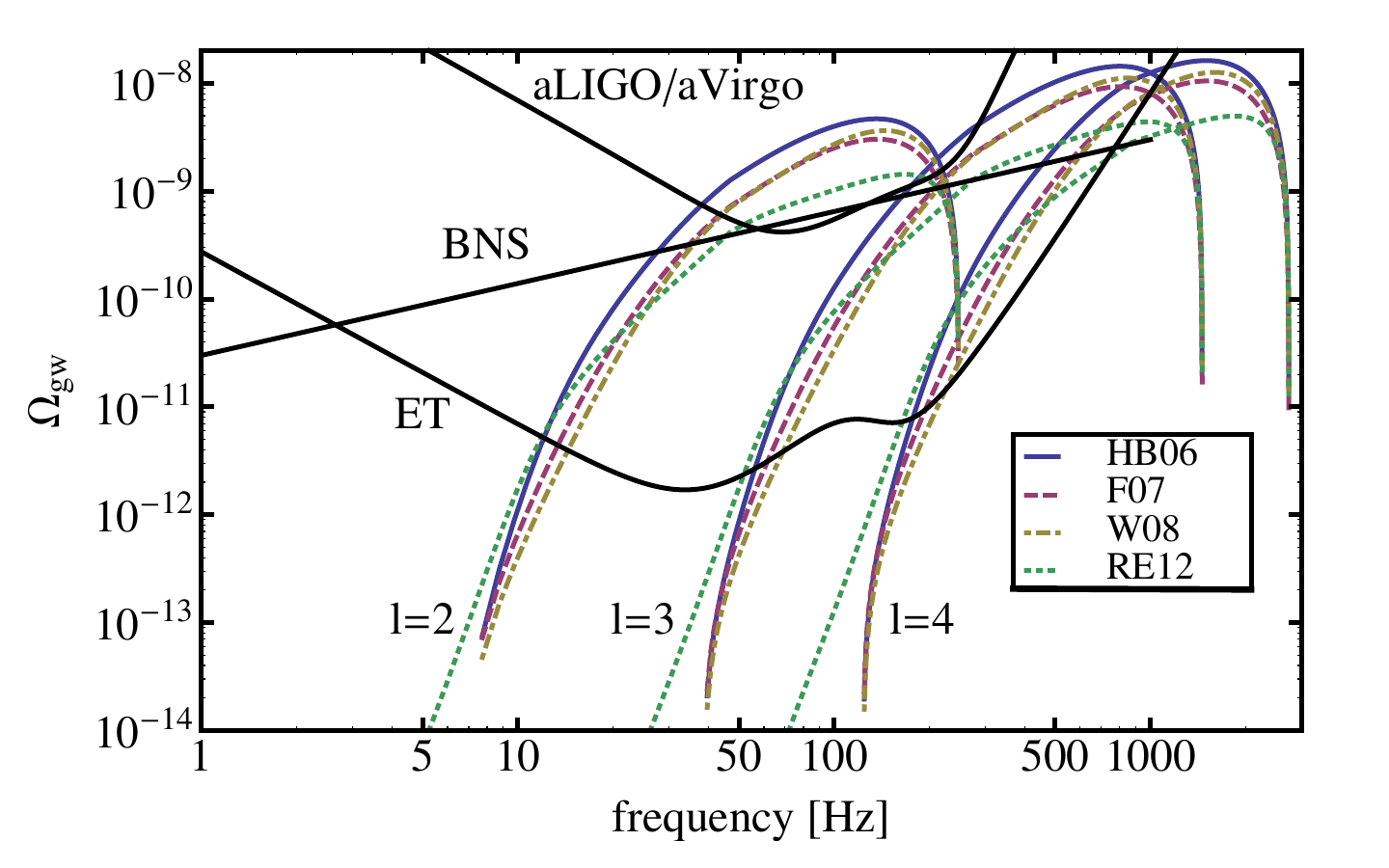}
\caption{Dimensionless energy density of the stochastic background for supernova-derived NSs, for four CSFR fits  (see Section II) and the $l=m=2,3,4$ $f$-modes, as a function of the observed frequency. Here 10\% of the NS population become $f$-mode unstable and 100\% of the rotational energy lost during the instability is due to GWs.}
\label{Omegahigh}
\end{center}
\end{figure}

 \begin{figure}[]
\begin{center}
\includegraphics[width=0.5\textwidth]{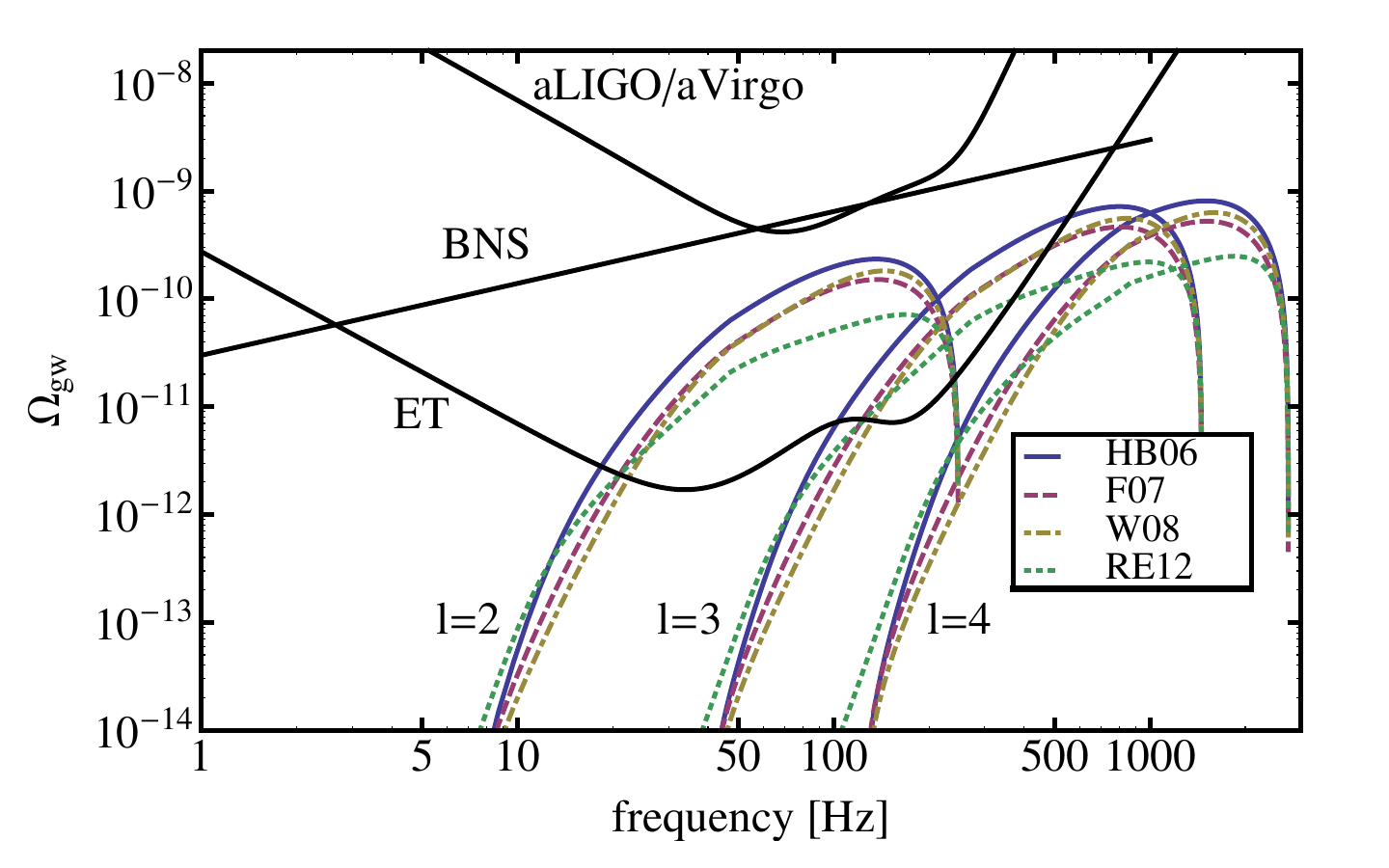}
\caption{Lower limit on the dimensionless energy density of the stochastic background for supernova-derived NSs, for four CSFR fits  (see Section II) and the $l=m=2,3,4$ $f$-modes, as a function of the observed frequency. Here 1\% of the NS population become $f$-mode unstable and 50\% of the rotational energy lost during the instability is due to GWs.}
\label{Omegalow}
\end{center}
\end{figure}

 \begin{figure}[]
\begin{center}
\includegraphics[width=0.5\textwidth]{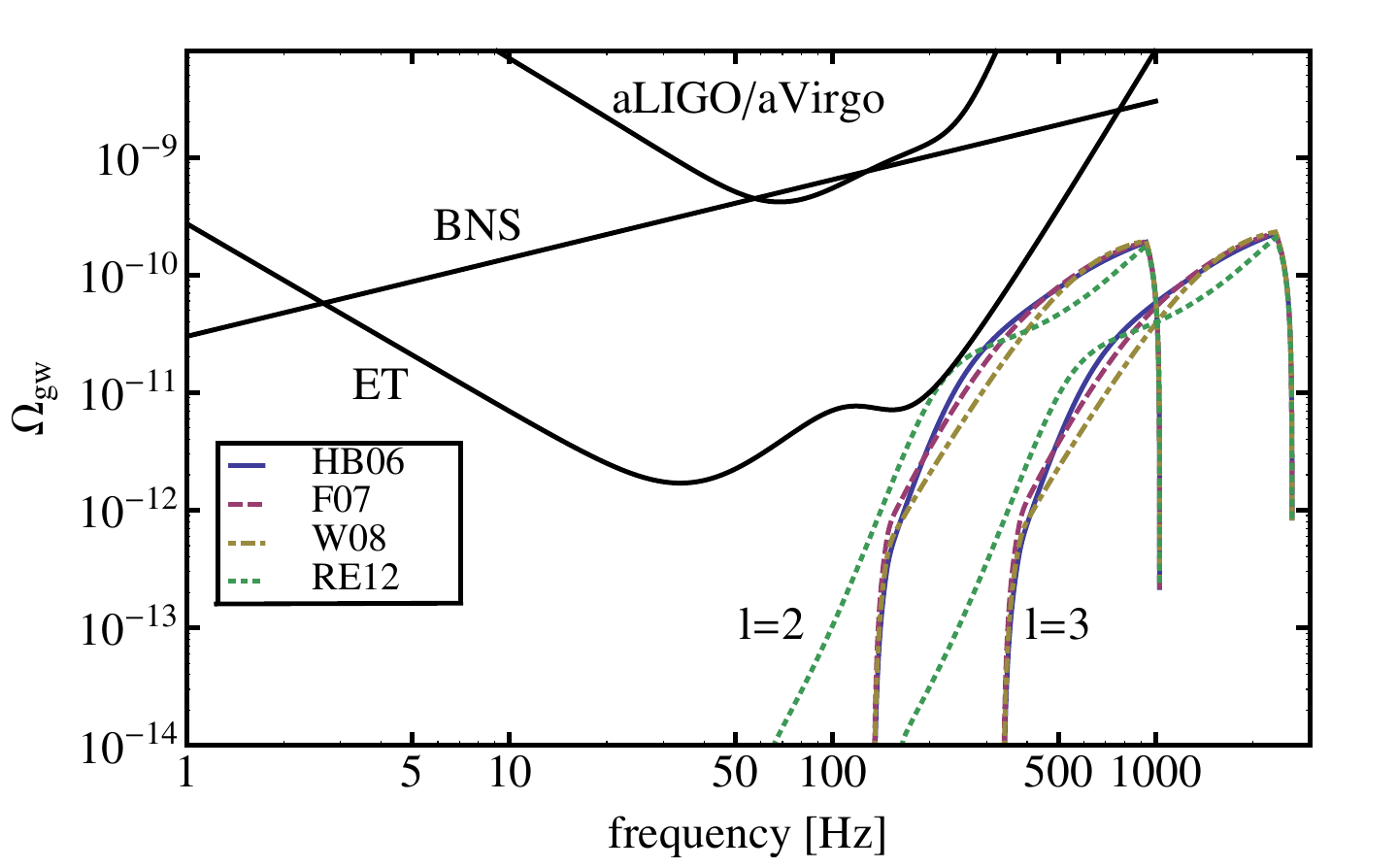}
\caption{Dimensionless energy density of the stochastic background for binary merger-derived NSs, for four CSFR fits  (see Section II) and the $l=m=2,3$ $f$-modes, as a function of the observed frequency.}
\label{Omegabin}
\end{center}
\end{figure}

\subsection{Duty cycle}

Another important quantity for an astrophysical SGWB is the duty cycle $D$, which classifies the background into three regimes: a continuous background ($D \gg1$) where the waveforms overlap to produce a background with Gaussian properties, shot noise ($D \ll1$) where the waveforms are separated by long stretches of silence, or popcorn noise ($D \sim1$) where the waveforms may overlap but Gaussian statistics are not obeyed. The duty cycle, in the observer frame, is defined as \citep{1674-4527-11-4-001}
\begin{equation} D=\int^\infty_0 \bar{\tau}(1+z')\frac{dR(z')}{dz'}\,dz'\,. \end{equation}
where $\bar{\tau}$ is the average time duration of the GW emission from a single source in the source frame, which dilates to  $\bar{\tau}(1+z)$ due to the cosmic expansion. 

For low magnetised ($B_p \leq 10^{11}$ G) supernova-derived NSs, the best estimate available for the evolution time $\bar{\tau}$ is 200 yr \citep{PhysRevD.87.084010}, and for merger-derived NSs the value is 24 hr \citetext{D.~Doneva, private communication}, giving $D \sim 10^{12}$ and $\sim 10^5$ respectively, thus we would expect to observe a continuous GW background in both cases. Taking into account a shorter evolution time for stars with a stronger magnetic field, then $\bar{\tau} \sim$ 10 yr \citep{PhysRevD.87.084010} and $\approx$ 20 min \citetext{D.~Doneva, private communication}, and $D \sim10^{10}$ and $\sim 10^3$, for supernova and merged-derived NSs with $B_p \sim 10^{12}$ G and $\sim10^{14}$ G, respectively. Thus we would still expect a continuous background.

Note that the time taken for the $f$-mode to saturate is a very small fraction of the total evolution time, so the duration of the GW emission is approximately equal to the total evolution time. The estimates for the evolution time that have been used depend upon the saturation energy of the mode, where a value of $\sim10^{-6} M_\odot c^2$ has been assumed for both NS classes. These estimates could be affected by more recent work, suggesting values as low as $10^{-10} M_\odot c^2$ for the saturation energy of supernova-derived NSs \citep{PnigourasKokkotasUnpublished, PnigourasKokkotas2015}. This, however, would only affect the duty cycle and would lengthen the evolution time, which increases the value of $D$.

\section{Detectability}

Having confirmed that we expect to observe a continuous GW background with Gaussian properties, we can employ the optimal detection strategy which involves cross-correlating measurements from two or more detectors with uncorrelated noise. This is required because the stochastic background can be confused with the intrinsic noise background of a single detector. Assuming the background to be isotropic, unpolarized, stationary, and Gaussian, the optimal signal to noise ratio (S/N) for cross-correlating two L-shaped interferometers during an observation time $T$ is given by \citep{PhysRevD.59.102001}
\begin{equation} \left( \frac{\text{S}}{\text{N}} \right)^2 = \frac{9H_0^4}{50\pi^4}T \int^\infty_0df\frac{\Gamma^2(\nu)\Omega^2_{\text{gw}} (\nu_o)}{\nu_o^6S^{(1)}_h(\nu_o)S^{(2)}_h(\nu_o)}\,,  \end{equation}
where $S_h^{(1)}(\nu_o)$ and $S^{(2)}_h(\nu_o)$ are the power spectral noise densities of the two detectors (from Table 1 of \cite{livrev}), and $\Gamma(\nu_o)$ is the overlap reduction function, characterizing the loss of sensitivity due to the separation and the relative orientation of the detectors.

\begin{table}[h]
\begin{center}
\begin{tabular}{lllll} \hline
         Pair & $l=m=2$ & $l=m=3$ & $l=m=4$ & \\ \hline
aL/aV &9.80 &1.53 &6.07$\times10^{-2}$   \\      
ET &1130  &193  &10.6	\\
\end{tabular}
\end{center}
\caption{Optimal signal-to-noise ratio obtained by correlating interferometer pairs for the $l=m=2,3,4$ $f$-modes and supernova-derived NSs with the HB06 CSFR  (see Section II). Here 10\% of the NS population becomes $f$-mode unstable and 100\% of the rotational energy lost during the instability is due to GWs.}
\label{}
\end{table}

\begin{table}[h]
\begin{center}
\begin{tabular}{lllll} \hline
         Pair& $l=m=2$ & $l=m=3$ & $l=m=4$ & \\ \hline
aL/aV &0.490 &7.66$\times10^{-2}$ &3.04$\times10^{-3}$   \\      
ET &56.5  &9.63 &0.530	\\
\end{tabular}
\end{center}
\caption{Lower limit on the optimal signal-to-noise ratio obtained by correlating interferometer pairs for the $l=m=2,3,4$ $f$-modes and supernova-derived NSs with the HB06 CSFR  (see Section II). Here 1\% of the NS population becomes $f$-mode unstable and 50\% of the rotational energy lost during the instability is due to GWs.}
\label{}
\end{table}

\begin{table}[h]
\begin{center}
\begin{tabular}{lllll} \hline
         Pair& $l=m=2$ & $l=m=3$ &  & \\ \hline
aL/aV &5.34$\times10^{-3}$ &4.31$\times10^{-5}$   \\      
ET &7.00$\times10^{-1}$  &1.64$\times10^{-2}$  & 	\\
\end{tabular}
\end{center}
\caption{Optimal signal-to-noise ratio obtained by correlating interferometer pairs for the $l=m=2,3$ $f$-modes and binary merger-derived NSs with the HB06 CSFR  (see Section II).}
\label{}
\end{table}

The optimal S/N has been evaluated for a pair of second and third generation GW detectors, aLIGO/aVirgo and ET respectively, for the two NS classes considered in this work, and for the HB06 CSFR fit which provides the highest rate of NS formation. We have assumed two co-located detectors with optimal orientation (i.e. $\Gamma(\nu_o)=1$), and an observation time of $T=3$ yr. The results, which represent an optimistic estimate of the S/N, are shown in Table I, II and III for supernova and merger-derived NSs respectively.

The results in Tables I$-$III agree with the detectability outcomes in Fig. $6-8$, and hence provide a consistency check on the previous results. 

\section{Conclusions}

In this paper, the integrated flux and dimensionless energy density of the SGWB produced by a population of hot, young, rapidly rotating NSs has been evaluated, which emit GWs during the spin-down phase associated with the $f$-mode instability. Two classes of NSs have been considered, those formed from supernovae, and supramassive NSs formed through the merger of binary NS systems. Four different parameterized fits to the expected evolution of the CSFR density have been extracted from \cite{0004-637X-651-1-142, Fardal11082007, Wilkins01042008, 0004-637X-744-2-95} in order to account for the uncertainty upon its determination. The $f$-mode frequencies and the instability windows have been extracted from \cite{PhysRevD.88.044052} and \citep{PhysRevD.92.104040} for supernova and merger derived NSs respectively, for a relativistic star with the WFF2 EoS in the Cowling approximation.

Changing the CSFR fit used in the calculations, which determines the GW event rate, does not change the detectability of a particular background, since the signal is mainly contributed by low redshift sources ($z<0.5$) where the rates of source formation (Fig. 2 and Fig. 3) are in close agreement. The fact that the RE12 fit extends to far higher redshifts than the other fits leads to a small increase in signal strength at low frequencies.

Changing the EoS used in the calculations would change the frequency of the GWs emitted, and also the total energy emitted via GWs, since the instability window would be affected. Although the GW spectrum would shift in frequency and/or flux, the position of the peak relative to the rest of the curve would not change, since it depends upon the CSFR. According to Fig. 2 and Fig. 11 of \cite{PhysRevD.88.044052} however, the realistic AkmalPR EoS \citep{AkmalEtAl1998} does not lead to significantly different unstable $f$-mode frequencies or instability windows compared to the WFF2 EoS. This, in addition to the dependence of the signal strength on the fraction of NSs which become $f$-mode unstable and the strength of the NS magnetic field, means that constraining the NS EoS using the $f$-mode SGWB is unlikely.

The dimensionless energy density $\Omega_{\text{gw}}$ in Fig. 6 for the background produced by supernova-derived NSs was found to peak at $\sim 10^{-9}$ and $\sim 10^{-8}$ for the $l=m=2$ and $l=m=3,4$ respectively, at frequencies of $\sim$ 200, 1000, 2000 Hz respectively. The $l=m=2$ background should be detectable with second generation interferometers. The lower estimate for the background in Fig. 7, assuming fewer NSs become $f$-mode unstable and not all the rotational energy lost is due to GW emission, peaks at an order of magnitude lower amplitude at the same frequencies. It should still be detectable with the ET interferometer, but is weaker than the BNS background over the entire frequency range. The background produced by supramassive NSs formed from binary mergers in Fig. 8 peaks at $\Omega_{\text{gw}} \sim 10^{-10}$ at frequencies $\sim$ 1000 and 2000 Hz for the $l=m=2$ and 3 $f$-mode respectively. It will not be detectable, even with ET.

The GW detections that should be made in the coming years with second generation detectors will provide fascinating new insights into the sources. The detection of the $f$-mode SGWB will improve our understanding of the astrophysical SGWB, whose shape needs to be understood in order for it to be disentangled from the primordial background.

\section*{Acknowledgements}

We gratefully acknowledge the support of the German Science Foundation (DFG) via SFB/TR7.

%
 \bibliographystyle{aa} 
 \bibliography{paper}
%
\end{document}